\documentclass[epj]{svjour}
\usepackage[latin2]{inputenc}
\usepackage{graphicx}
\usepackage{amsmath}

\begin{document}
\title{Divergent estimation error in portfolio optimization and in linear regression}
\author{I. Kondor\inst{1}\inst{2}\inst{4} \and I. Varga-Haszonits\inst{2}\inst{3}\inst{5}}
\institute{Collegium Budapest - Institute for Advanced Study, Szentháromság u. 2, H-1014 Budapest, Hungary\and Department of Physics of Complex Systems, Eötvös University, Pázmány Péter sétány 1/A, H-1117 Budapest, Hungary\and Analytics Department of Fixed Income Division, Morgan Stanley, Budapest, Hungary\and email: kondor@colbud.hu\and email: Istvan.Varga-Haszonits@morganstanley.com}
\abstract{
The problem of estimation error in portfolio optimization is discussed, in the limit where the portfolio size $N$ and the sample size $T$ go to infinity such that their ratio is fixed. The estimation error strongly depends on the ratio $N/T$ and diverges for a critical value of this parameter. This divergence is the manifestation of an algorithmic phase transition, it is accompanied by a number of critical phenomena, and displays universality. As the structure of a large number of multidimensional regression and modelling problems is very similar to portfolio optimization, the scope of the above observations extends far beyond finance, and covers a large number of problems in operations research, machine learning, bioinformatics, medical science, economics, and technology.
}
\maketitle

\section{Introduction}
If we do not have sufficient information we are not able to take an optimal decision, nor to build a reliable model. The purpose of this paper is to turn this trivial remark into a quantitative statement. In contrast to the usual set-up of classical statistics where the dimension $N$ of the problem is considered fixed (and possibly small) while the sample size $T$ is assumed to be very large, we are considering a situation which is much more realistic in the context of complex systems: we assume that $N$ is large, and $T$ is limited, at most commensurate with $N$, that is we consider the 'thermodynamic limit' where $N/T$ is fixed while both $N$ and $T$ go to infinity. It will be seen that the estimation error strongly depends on the ratio $N/T$, so strongly indeed, that at a critical value of this ratio the estimation error actually diverges. This divergence has been recognized \cite{kondor_1} as the manifestation of an algorithmic phase transition \cite{Mezard}. As such, it is accompanied by a number of critical phenomena. The critical index of the divergence of the estimation error is universal: it does not depend on the details of the various models considered, such as the covariance structure of the market \cite{kondor_3}, the risk measure used \cite{kondor_1},\cite{ciliberti_1},\cite{ciliberti_2}, or the nature of the underlying stochastic process \cite{hasszan}. All these features have been illustrated on the example of portfolio optimization, but, in fact, there is nothing special about the financial aspect of the problem: it will arise in any other optimization where the parameters of the cost function have to be determined from empirical observations and we have a limited amount of information available. The obvious relationship between quadratic optimization and linear regression allows us to extend these conclusions also to the regression problem. 

The paper is organized as follows. In Sec. \ref{sec:simpleproblem} we recapitulate the most important results obtained for the quadratic optimization problem, while in Sec. \ref{sec:widercontext} we enlarge the scope of our study, display the obvious connections between optimization and regression, and cast them into a form that allows to formulate them as a problem in statistical mechanics. This way, a whole arsenal of powerful methods (scaling, phase transition concepts such as universality, random matrices, replicas) become available for the treatment of the ubiquitous problem of estimation error. The paper ends on a short summary.

\section{A simple quadratic optimization problem}
\label{sec:simpleproblem}

Let us consider the simplest portfolio optimization problem:
\begin{gather}
\min_{\left\{w_i\right\}}\sigma_p^2=\min_{\left\{w_i\right\}}\sum_{i,j}w_i\sigma_{ij}w_j,\label{eq:globminvar}
\end{gather}
subject to
\begin{gather}
\sum_iw_i=1
\end{gather}
The solution is the minimal variance portfolio with optimal weights given by
\begin{gather}
w_i^*=\frac{\sum_j\sigma_{ij}^{-1}}{\sum_{jk}\sigma_{jk}^{-1}}.
\end{gather}
Several remarks are in order:
\begin{itemize}
\item The use of the variance as a risk measure assumes that the underlying stochastic process is Gaussian, or at least has a similarly narrow distribution. It is well known that financial fluctuations often have fat tailed distributions for which the variance is not an adequate characteristic of risk. Most of what we are going to describe below is valid, mutatis mutandis, also for a number of other convex risk measures. A remark concerning other risk measures will be made at an appropriate point below.
\item Normally one does not seek the global minimal variance portfolio, but a portfolio that has minimal variance {\it given} a certain value of the expected return. The constraint on expected return has been dropped here, for simplicity. It should be noted, however, that the optimization task as spelled out here appears in an index tracking context. 
\item The constraint on the portfolio weights only stipulates that they sum to unity, but we do not assume that they are positive. That is we allow unlimited short positions. The case of excluded short selling will be mentioned later. 
\item The problem above will be regarded  more as a representative of a wide class of convex optimization problems than a realistic problem in finance. 
\end{itemize}
The covariance matrix in Eq. \eqref{eq:globminvar} has to be determined from observations on the market. For a portfolio of size N we need $O(N^2)$ elements. Time series of length $T$ for $N$ assets contain $N\times T$ input data. For the statistical estimate to be reliable we obviously need $NT\gg N^2$, that is $T\gg N$. Real life banking portfolios are large, they may contain several hundred elements, whereas the available time series are always limited. The choice of the length and frequency of the time series is dictated by considerations of stationarity and transaction costs, respectively. In practice $T$ is never longer than $T=1000$ (corresponding to four years worth of daily data), and often much shorter. (E.g. the EWMA method advocated by Riskmetrics \cite{riskmetrics} starts to cut off around a three-months time horizon.) Therefore, the inequality $N/T\ll 1$ almost never holds in practice, and we have to live with the consequences of this information deficit. In particular, our empirical covariance matrix and the portfolio weights obtained from it will contain a lot of estimation error, and the resulting portfolio will be suboptimal.

This problem is, of course, not new: economists have been struggling with this 'curse of dimensions' ever since the appearence of rational portfolio choice \cite{Markowitz}. Since the root of the problem is lack of sufficient information, the remedy is to inject external information into the estimate. This means imposing some structure on $\sigma_{ij}$. This introduces bias, but the beneficial effect of noise reduction may compensate for this. Elton and Gruber \cite{for_1} give a comprehensive review of the main methods. Our focus in this paper is the quantitative characterization of the problem, not the analysis of its remedies.

In the course of the analysis of the above optimization problem we have been using both analytic and numerical methods. On the analytical side we have applied methods borrowed from statistical physics, such as random matrix theory, phase transition theory, replicas, etc. On the numerical side we have used simulated data, so as to have full control over the underlying stochastic process and avoid problems related to non-stationarity and other imperfections, unrelated to estimation error. For simplicity, we have mostly used iid normal variables, but have considered other, more realistic underlying processes as well.

For such simple underlying processes the true risk measure can be calculated exactly. To construct the {\it empirical} risk measure
\begin{gather}
\sigma^{(1)}_{ij}=\frac{1}{T}\sum_{t=1}^Tx_{it}x_{jt}
\end{gather}
we generate long time series of the returns $x$, and cut out segments of length $T$ from them, as if making observations on the market. From these ,,observations'' we construct the empirical risk measure and optimize our portfolio under it.

The ratio $q_0$ of the empirical and the exact risk measure is a measure of the estimation error due to noise:
\begin{gather}
q_0^2=\frac{\sum_{ij}{w_i^{(1)}}^*\sigma_{ij}^{(0)}{w_j^{(1)}}^*}{\sum_{ij}{w_i^{(0)}}^*\sigma_{ij}^{(0)}{w_j^{(0)}}^*}\label{eq:q0def}
\end{gather}
Here $\sigma_{ij}^{(0)}$ is the exact covariance matrix, and $w_i^{(0)}$ are the portfolio weights optimized under $\sigma_{ij}^{(0)}$; $\sigma_{ij}^{(1)}$ is the empirical covariance matrix and $w_i^{(1)}$ are the optimal weights corresponding to $\sigma_{ij}^{(1)}$. The quantity defined in \eqref{eq:q0def} plays a role of central importance in our considerations.

\begin{figure}
\includegraphics*[width=9cm]{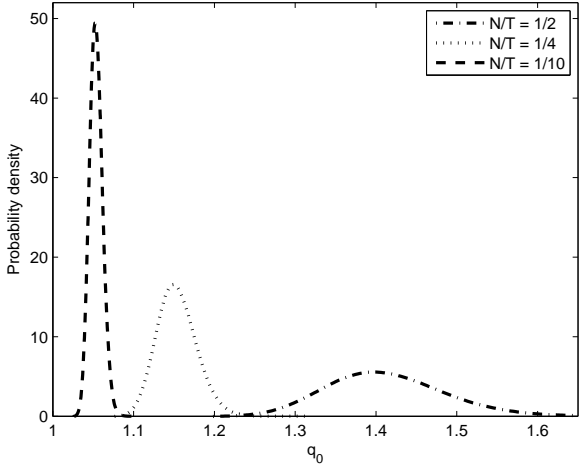}
\caption{\label{fig:q0vsNT}Distribution of $q_0$ over the samples computed by Monte Carlo simulations. The number of variables was $N=200$, and 50000 samples were generated to produce the histograms.}
\end{figure}

As defined in \eqref{eq:q0def}, $q_0$ depends on the optimal weights corresponding to a given sample, so it is a random variable, fluctuating from sample to sample. The weights of the optimal portfolio also fluctuate. The distribution of $q_0$ over the samples is shown in Fig. \ref{fig:q0vsNT}.
Since the weights in the numerator of \eqref{eq:q0def} are optimal under $\sigma_{ij}^{(1)}$, not $\sigma_{ij}^{(0)}$,  $q_0$ is always larger or equal to one. Its deviation from unity is the measure of estimation error. As seen in Fig. \ref{fig:q0vsNT}, the average of $q_0$ increases with $N/T$ . 
	
\begin{figure}
\includegraphics*[width=9cm]{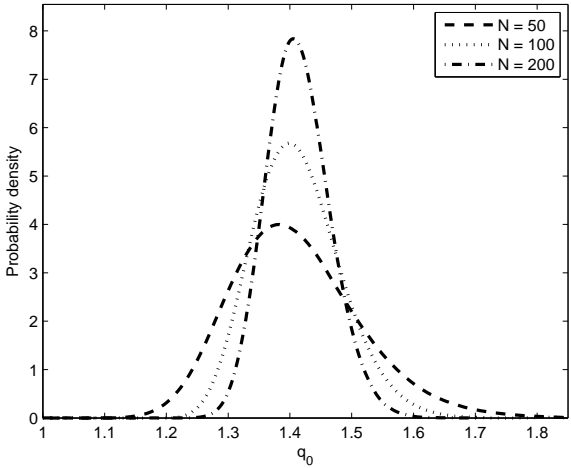}
\caption{\label{fig:q0vsN} Dependence of the distribution of $q_0$ on the number of variables $N$ with $N/T$ fixed at 1/2. 50000 samples were generated to produce the histograms.}
\end{figure}

The average of $q_0$ can, in fact, be calculated exactly by various methods. The first calculation, based on random matrix theory, was given in \cite{lengyelek}.
The result is the simple formula:
\begin{gather}
q_0=\frac{1}{\sqrt{1-N/T}}.\label{eq:q0mean}
\end{gather}
Eq. \eqref{eq:q0mean} shows that the average estimation error diverges at the critical value of the ratio $N/T=1$. The divergence of $q_0$ is the manifestation of an algorithmic phase transition, related to the appearence of zero modes in the empirical covariance matrix in the limit $N/T\to1$. We note that while for a fixed value of $N/T$ the width of the distribution of $q_0$ tends to zero with increasing $N$ (i.e. $q_0$ is a self-averaging quantity, see Fig. \ref{fig:q0vsN}.), for fixed $N$ and $N/T$ going to its critical value, the width diverges even stronger than the average.
	
For iid variables the true covariance matrix is proportional to the unit matrix, and the true portfolio weights are all the same $1/N$, therefore \eqref{eq:q0def} works out to be
\begin{gather}
q_0^2=N\sum_{i}{w_i^{(1)}}^2.
\end{gather}
The large fluctuations of $q_0$ are thus related to the large fluctuations in the length of the solution vector. The individual components of this vector, the portfolio weights, show strong fluctuations already for relatively small values of $N/T$, such as $1/5$ or $1/3$. It is easy to show that the standard deviation of the weights diverges at $N/T=1$ with the same index $-1/2$ as $q_0$. This means that the optimization hardly determines the weights even far from the critical point! 

It is evident that the above divergences are related to the unrealistic feature of our optimization task: the budget constraint allows arbitrarily large short positions, as long as they are compensated by similarly large long positions. Accordingly, when the cost function becomes flat in the limit $N/T\to1$, the solution vector may display arbitrarily large fluctuations. A ban on short selling, that is, demanding that all the weights be positive, will eliminate these fluctuations, which explains the observation that a ban on short selling acts as a regulator \cite{jagannathan}.We might then believe that via the constraint $w_i\ge0$ we can save the optimization task. This is not so: the critical fluctuations will indeed be eliminated, but as we go close to the value $N/T=1$, and even more as we go into the region $N/T>1$, we find that an increasing number of the weights get stuck to the boundaries of the allowed region, that is a larger and larger number of weights becomes zero. This phenomenon of spontaneous reduction of portfolio size is well known \cite{book}, and is easy to understand on the basis of our geometric picture: with increasing $N/T$ an ever increasing number of zero modes appear, the cost function becomes flat in more and more directions, and the solution would like to run away to infinity along these soft directions. It is, however, prevented from this by the positivity constraints imposed on the weights, so it gets glued to the coordinate planes representing these constraints. Moreover, the fluctuations from sample to sample will now mean that the solution is randomly jumping about the coordinate planes \cite{gulyas}, therefore the solution of the problem will not reflect any stable, objective structure of the cost function.  Clearly, any other set of constraints that makes the domain of the optimization bounded will have a similar effect \cite{gulyas}.

So far we have been considering the oversimplified case of iid variables. What happens if we relax this condition? Experimenting with various market models (one-factor, market plus sectors, positive and negative covariances, etc.) we have come to the conclusion that the main features of the task do not change, in particular the exponent of $q_0$ remains the same \cite{kondor_1}. This is a manifestation of universality: the critical index is invariant to the microscopic details of the cost function. In fact, universality goes much farther than this. It has been shown that the exponent of $q_0$ is also largely independent of the nature of the underlying process and remains the same for fat tailed distributions with a tail index $\alpha>2$ \cite{student}, moreover, it remains the same even for a special type of nonstationary process, the Constant Conditional Correlation GARCH process \cite{hasszan}.The phase transition concept can thus incorporate a number of different aspects of the problem (independence of the risk measures and the underlying process, reduction of diversification for banned short selling) into a single coherent picture.

\section{A wider context}
\label{sec:widercontext}

We have seen that the estimation error of the optimal portfolio risk diverges as the number of observations approaches the number of stocks in the portfolio. Obviously, this estimation instability is not restricted to portfolio optimization, but occurs in a very wide range of applications also beyond finance. In fact, whenever a phenomenon is influenced by a large number of factors, but we have a limited amount of information about this dependence, we have to expect that the estimation error will diverge and fluctuations over the samples will be huge. To show this, we first consider a slightly generalized version of the optimization problem in the previous Section by introducing a set of linear constraints, then point out that linear regression can also be cast into this form. So the instability due to information deficit is present not only in optimization, but also in model building. 

Let us then consider the following class of constrained quadratic optimization problems:
\begin{gather}
\min_{\{w_i\}}\sum_{i,j=1}^N\sigma_{ij}w_iw_j+\sum_{i=1}^Nh_iw_i\notag\\
\sum_{j=1}^NA_{kj}w_j=b_k\label{eq:quadropt}\\
k=1,2,...,K\notag
\end{gather}
where we assume that the number $K$ of constraints is smaller than $N$, $\mathrm{rank}(\mathbf{A})=K$, and $\sigma$ is a symmetric, positive definite matrix. Important special cases of this general framework are the following:
\begin{enumerate}
\item Portfolio optimization: $\sigma_{ij}=\mathrm{Cov}(x_i,x_j)$ ($i,j=1,...,N$) where $x_i$ is the return on asset $i$ and the variables $w_i$ are the portfolio weights. The coefficients $h_i$ are all zero and the constraints may vary according to the kind of optimum we are looking for. Two particular cases deserve special mention:
	\begin{enumerate}
	\item Global minimum variance portfolio: $K=1$ and $A_{1j}=1$ for all $j=1,2,...,N$, which is precisely the simple case considered in the previous Section.
	\item Mean-variance portfolio optimization: $K=2$, $A_{1j}=1$ and $a_{2j}=\mu_j$ for all $j = 1,2,...,N$, where $\mu_j$ is the expected return of asset $j$, the standard textbook example of portfolio optimization.
	\end{enumerate}
\item Linear regression: consider the regression equation:
\begin{gather}
y=w_0+\sum_{i=1}^Nw_ix_i+\epsilon.
\end{gather}
The task is to minimize the variance of the deviation $\epsilon$ over the regression coefficients $w_i$, so we have a quadratic optimization problem with $\sigma_{ij}=\mathrm{Cov}(x_i,x_j)$ ($i,j = 1,...,N$),  $K=0$ (no constraints), and $h_i=Cov(x_i,y)$.
\end{enumerate}

It is well-known that problems conforming to \eqref{eq:quadropt} can be transformed into several equivalent forms with fewer or more constraints via linear transformations of the variables and parameters, and/or eliminating/introducing new variables (Lagrange multipliers). Thus, any quadratic optimization problem of the form \eqref{eq:quadropt} can be rephrased either as a portfolio optimization, or as a linear regression problem, which allows results obtained in one field to be transferred to the other \cite{Kempf}. On the other hand, we can always regard the cost function in \eqref{eq:quadropt} as a simple quadratic Hamiltonian, so the task set forth in \eqref{eq:quadropt} is nothing but finding the ground state energy of that Hamiltonian. Introducing a fictitious temperature we can then turn the optimization problem into one in statistical mechanics, an idea that provided the key to the method of simulated annealing \cite{Kirkpatrick}. The partition function of this statistical mechanics problem is:
\begin{gather}
Z=\int\exp\left[-\beta\sum_{ij}\sigma_{ij}w_iw_j-\beta\sum_ih_iw_i\right]\prod_idw_i
\end{gather}
The solution to the original problem can then be recovered in the zero temperature ($\beta\to\infty$) limit.

Just as in the case of finance, in most other disciplines the parameters $\sigma_{ij}$, $h_i$ and $A_{kj}$ are partially or fully estimated from empirical samples, hence they fluctuate from sample to sample. Therefore, we face a situation again where the estimated minimum of the objective function is actually suboptimal. The amount of measurement noise can be measured by the same $q_0$ as we used before.

Let us denote the cost function (Hamiltonian) by $\mathcal{H}(w_i; \alpha_i)$ where we optimize over the variables $w_i$. The vector $\alpha_i$ includes all fixed parameters of the problem ($\sigma_ij$, $h_i$ and $A_kj$). In theory, the optimum can be expressed as a function of the fixed parameters. In practice, however, the real value $\alpha^{(0)}_i$ of some or all of these parameters are unknown (and so are the real optimal weights ${w_i^{(0)}}^*$), all we can have is a sample based estimator $\alpha^{(1)}_i$, and the corresponding estimated optimum will be ${w_i^{(1)}}^*$. Then the measure of estimation error can generally be defined as:
\begin{gather}
q_0^2=\frac{\mathcal{H}\left({w_i^{(1)}}^*;\alpha^{(0)}_i)\right)}{\mathcal{H}\left({w_i^{(0)}}^*;\alpha^{(0)}_i\right)}.\label{eq:q0gen}
\end{gather}
It is easy to see that the quantity defined here is equivalent to the $q_0$ introduced in the first part of this paper.

Let us interpret this result for the problem of linear regression. In this case, the cost function is the mean squared error of the linear model, that is, the variance of the difference between the observed value and the predicted value of the dependent variable. The denominator of \eqref{eq:q0gen} is the theoretical minimum of this variance characterizing the line/plane/hyperplane that best fits the true model (i.e. the equation with the true parameter values). On the other hand, the quantity in the numerator is the expected value of the mean squared error of the estimated model. In other words, it characterizes the prediction error of the model with the estimated (and hence suboptimal) parameter values. This latter quantity is obviously always larger than the former one, so $q_0-1$ represents the relative increase in the prediction error due to sampling noise.

In the financial context, the properties of $q_0$ have been extensively investigated 
\cite{kondor_1}, \cite{kondor_2}, \cite{kondor_3}. An analytic calculation of the mean of $q_0$ has been carried out by \cite{lengyelek} using results borrowed from random matrix theory. On the other hand, computing the expectation of $q_0$ means averaging over several samples generated by the same underlying process. This is completely analogous to quenched averaging in spin glasses: one has to compute the expected value of the logarithm of the partition function. This computation can be performed using the replica trick. Indeed, the replica method has been employed successfully to investigate the instability of $q_0$ for different risk measures \cite{ciliberti_1}, \cite{ciliberti_2}. We have rederived the average of  $q_0$ for the variance as risk measure also for correlated Gaussian underlying processes, and confirmed the numerical finding in \cite{kondor_2} that asymptotically the behaviour of  $q_0$ is the same as for the iid normal case considered in \cite{lengyelek}. The calculation of the higher moments of $q_0$ is underway \cite{hasszan_2}.

\section{Summary}

We have addressed the problem of estimation error in the context of portfolio optimization. Statistical methods work perfectly when the portfolio size is small and the sample large. Banking portfolios are almost never like that, the 'thermodynamic limit', where the portfolio size and the sample size go to infinity such that there is a fixed ratio between them, often represents the actual situation better. We found that at a critical value of this ratio the sampling error diverges. This is an algorithmic phase transition, displaying universal power law-like divergences. We have seen that this phase transition picture is more than a somewhat fancy description of the well known fact that the empirical covariance matrix develops zero eigenvalues for short time series: the phase transition concept encompasses a number of seemingly diverse phenomena and organizes them into a coherent picture, and it also encourages one to take over the extremly powerful methods of statistical mechanics to the optimization problem. Via the relationship between quadratic optimization and linear regression all these results can be extended also to the latter.
	
Complex systems depend on a large number of parameters, they are intrinsicly high dimensional, by definition. To collect sufficient data about them is often prohibitively expensive or practically impossible. Therefore one typically has to face an information deficit catastrophe similar to the one described above and the estimation error and sample to sample fluctuations will be huge. The description and modeling of such systems may benefit from the link statistical mechanics - optimization - regression described in this paper.

\section*{Acknowledgement}
This work has been supported by
the "Cooperative Center for Communication Networks Data Analysis"', a NAP project
sponsored by the National Office of Research and Technology under grant No. KCKHA005.

\end{document}